\documentclass[
reprint,
nofootinbib,
 amsmath,amssymb,
 aps,
]{revtex4-2}

\usepackage{graphicx}
\usepackage{dcolumn}
\usepackage{bm}\usepackage{graphicx}
\usepackage{textcomp}
\usepackage{color}
\usepackage{dsfont}
\usepackage{siunitx}
\usepackage{mathtools}
\usepackage{physics}
\usepackage{upgreek}
\usepackage{float}
\usepackage{tikz}
\usepackage{xcolor}
\usepackage{txfonts}
\usepackage{comment}
\usepackage[colorlinks=true, linkcolor=black, citecolor=black, urlcolor=blue]{hyperref}
\usepackage[normalem]{ulem}

\definecolor{LDA}{RGB}{184, 66, 3}
\definecolor{LDA03}{RGB}{240, 120, 24}
\definecolor{LDA05}{RGB}{254, 187, 71}
\definecolor{BEC}{RGB}{34, 139, 34}
\definecolor{fig31}{RGB}{157, 185, 188}
\definecolor{fig32}{RGB}{0, 0,0}
\definecolor{fig4a}{RGB}{64, 64,64}
\definecolor{stateup}{RGB}{251,110,78}
\definecolor{statedown}{RGB}{30, 115,220}
\definecolor{photon_cut}{RGB}{204, 178,0}
\definecolor{arrow_kplus}{RGB}{255, 0,0}
\definecolor{arrow_pauliassisted}{RGB}{0, 159,29}
\definecolor{arrow_pauliblocked}{RGB}{107, 23,199}
\DeclareSIUnit{\Gauss}{G}
\DeclareSIUnit{\rad}{rad}

\emergencystretch=\maxdimen
\begin{document}

\preprint{APS/123-QED}

\title{Fermi-pressure-assisted superradiant transition with a mesoscopic Fermi gas in a cavity}

\author{Francesca Orsi}
\author{Ekaterina Fedotova}
\author{Rohit Prasad Bhatt}
\author{Mae Eichenberger}
\author{Léa Dubois}
\author{Jean-Philippe Brantut$^{\ast}$}

\affiliation{
Institute of Physics and Center for Quantum Science and Engineering, Ecole Polytechnique Fédérale de Lausanne (EPFL), Lausanne, Switzerland
}

\begin{abstract}
We study the superradiant phase transition of a mesoscopic Fermi gas comprising between a few tens and a few thousand $^6$Li atoms in a high-finesse cavity across a wide range of densities. We observe a non-monotonic variation of the superradiant threshold as a function of density, with a minimum reached when the Fermi and recoil wavevectors are comparable. The minimum corresponds to a crossover between Fermi pressure-assisted ordering and Pauli blocking of photon scattering, in good agreement with theory. This interpretation is confirmed by a study of the atom-number dependence of the ordering threshold and photon number scaling. Our system opens the perspective of studying few-fermion systems with strong and coherent light-matter coupling.  
\end{abstract}

\maketitle

\let\thefootnote\relax\footnotetext{$^\ast$ Contact author: jean-philippe.brantut@epfl.ch}

The interplay of Fermi statistics with interactions is at the heart of our understanding of the microscopic world, from the stability of matter to the chemical properties or optical spectrum of elements. In a system of degenerate fermions subjected to an external drive, scattering processes can be suppressed by phase-space occupation constraints imposed by the Pauli principle, like in neutron scattering in solids or liquid Helium three \cite{nozieres_pines_1989}. 
In quantum gases, a paradigmatic example is the suppression of light scattering in a quantum-degenerate Fermi gas where photon recoil is Pauli blocked, which was predicted thirty years ago \cite{Busch_1998,Shuve_2010,PhysRevA.63.041601}, and only recently observed experimentally \cite{doi:10.1126/science.abi6153, 
doi:10.1126/science.abh3483, Jannin_2022, doi:10.1126/science.abh3470}. 

\begin{figure}[t]
\includegraphics{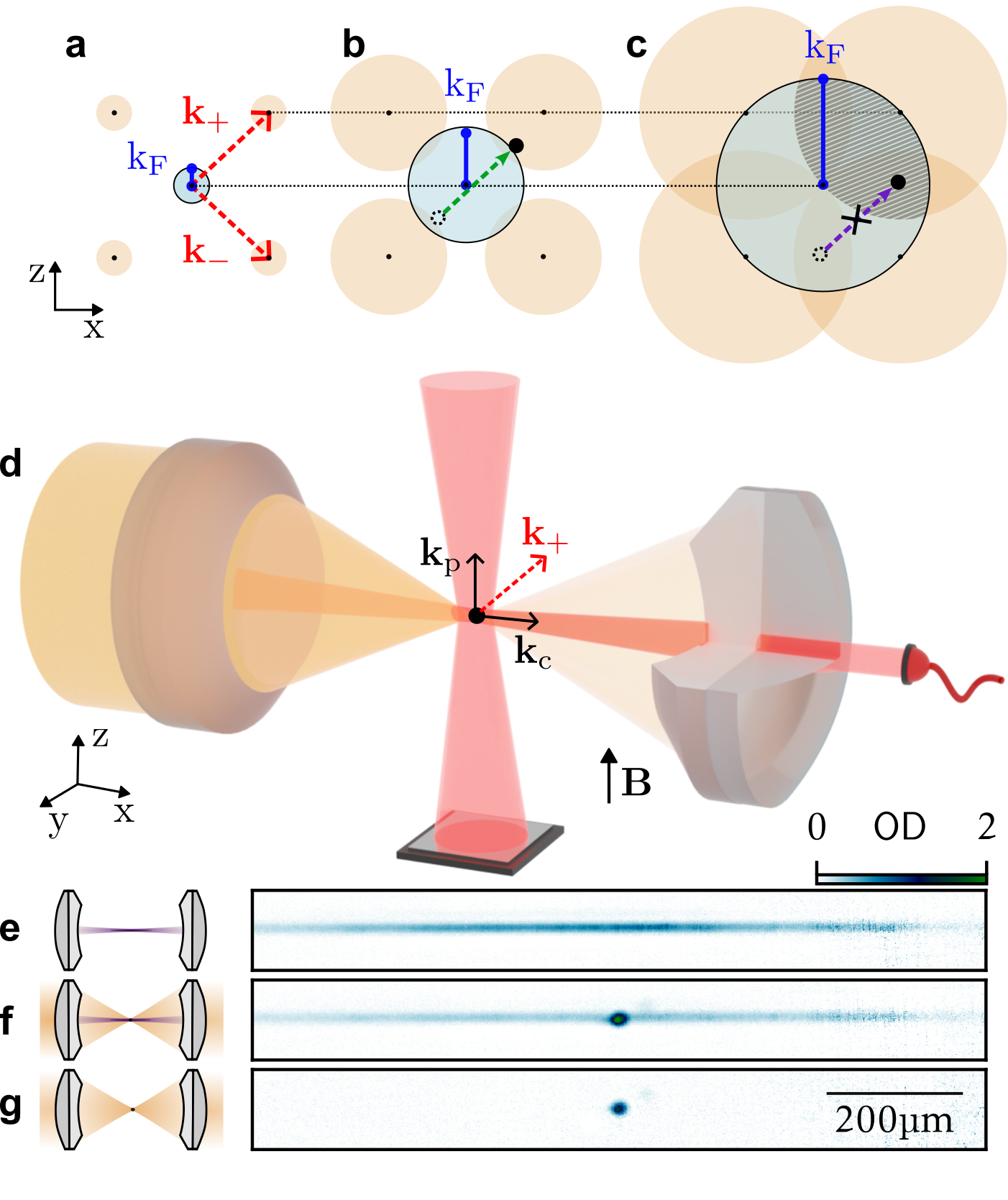}
\caption{\label{fig:fig1} Concept of the experiment.      (a) Recoil processes involved in the scattering from a pump laser into a cavity and vice versa at wavevectors $\pm\mathbf{k}_{\pm}$ (\raisebox{0.5ex}{\protect\tikz{\protect\draw[arrow_kplus, densely dashed, line width=0.4mm] (0,.8ex)--++(.3,0);}}), compared with the radius $\mathrm{k}_{\mathrm{F}}$ of the Fermi surface, at low (a), intermediate (b), and high density (c). (\raisebox{0.5ex}{\protect\tikz{\protect\draw[arrow_pauliassisted,densely dashed, line width=0.4mm] (0,.8ex)--++(.3,0);}}) shows an example process at $\mathbf{k_+}$ favored by the existence of the Fermi sea, and (\raisebox{0.5ex}{\protect\tikz{\protect\draw[arrow_pauliblocked,densely dashed, line width=0.4mm] (0,.8ex)--++(.3,0);}}) and example of Pauli-blocked $\mathbf{k_+}$ process. In (c), these processes correspond to the overlap between the Fermi sea (blue) and the Fermi sea displaced by $\pm\mathbf{k}_{\pm}$ (orange). (d) Cavity-microscope system comprising two mirrors contacted to aspherical lenses allowing to focus a tweezer trap (orange) onto an atomic cloud at the center of the cavity. The cavity mode with wavevector $\mathbf{k_c}$ (red) shares the same optical axis as the tweezer trap beam. A retro-reflected transverse pump beam with wavevector $\mathbf{k}_{\mathrm{p}} \bot \mathbf{k}_{\mathrm{c}}$ (red) drives the atoms. Photons leaking through one of the cavity mirrors are detected on a single-photon counter. (e-g) Absorption images of atoms in state $\ket{\downarrow}$ after evaporation in the cavity-dipole trap alone (purple, not shown in panel (d)), (e), in the cavity-dipole trap superimposed with the tweezer trap (f), and after releasing the cavity dipole trap, yielding about $\mathrm{N}= 4000$ atoms in the tweezer (g).}
\end{figure}

The setting of cavity quantum-electrodynamics (QED) provides the cleanest platform to investigate the interplay between light-matter interactions and atomic motion. Compared with free-space scattering, it isolates a single mode of the electromagnetic field, which can then be monitored and controlled with exquisite precision \cite{Tanji-Suzuki:2011ac, Haroche:2013aa, Reiserer:2015aa}. Most importantly, the cavity feeds the light scattered by an atom back onto its emitter, leading to strong nonlinearities capable of driving a gas into a distinct, superradiant phase of matter \cite{Black:2003aa,Baumann_2010,Klinder:2015aa, Koll_r_2017, doi:10.1126/science.abd4385, Helson_2023,Ho_2025}, observed recently for degenerate Fermi gases \cite{doi:10.1126/science.abd4385,Helson_2023}. It was found that, in a deep one-dimensional optical lattice, the atom-number scaling of the superradiance threshold exhibits signatures of Pauli exclusion, with a reduced power-law exponent in a degenerate gas compared with a thermal gas \cite{doi:10.1126/science.abd4385}. For interacting systems, the threshold was observed to increase as the system crosses over from a Bose-Einstein condensate (BEC) of molecules to a Bardeen-Cooper-Schrieffer superfluid, suggesting a role for Fermi statistics in hindering light scattering processes \cite{Helson_2023}. However so far Fermi statistics was only observed to suppress superradiance, in contrast with earlier predictions suggesting enhancement compared with a Bose-Einstein condensate in certain parameter regimes \cite{PhysRevLett.112.143002,PhysRevLett.112.143003,Chen_2014,ortuñogonzalez2025paulicrystalsuperradiance}.

In this letter, we systematically study the role of Pauli blocking in the cavity-induced superradiant transition in a three-dimensional mesoscopic Fermi gas using a novel cavity-microscope system combining a high-finesse cavity with a micrometer-scale tweezer trap. The setup allows for variations of the atomic density over two orders of magnitudes at fixed atom number, yielding a Fermi sea with varying radius. We find a non-monotonic threshold with a minimum when the recoil approaches twice the Fermi wavevector, in agreement with theory, a striking manifestation of the role of Fermi statistics in cavity superradiance. Our setup also allows to observe the superradiant threshold with varying atom number between a few tens up to thousands, an uncharted regime between experiments with few individual, thermal atoms in tweezers \cite{yan:2023aa, Ho_2025} and the regime of bulk, macroscopic quantum gases \cite{Black:2003aa,Baumann_2010,Klinder:2015aa, Koll_r_2017, doi:10.1126/science.abd4385, Helson_2023,Ho_2025}.

The cavity-induced superradiant transition originates from photon scattering between the cavity mode and a pump beam, when the frequency of the drive and the cavity mode resonance are far detuned from the atomic transition. Scattering from the pump into the cavity yields four discrete recoil momenta determined by the laser wavelength and the geometry, illustrated in Figure \ref{fig:fig1}a. At low densities where atoms only populate low momenta, the associated energy cost is the photon recoil energy, and superradiance is similar to that in Bose-Einstein condensates \cite{Baumann_2010}. When density increases, Fermi pressure forces atoms to populate higher momentum states. As a result, photon scattering can occur at a reduced energy cost by going across the Fermi sea, as shown in Fig. \ref{fig:fig1}b. 
This effect relies on the {\it width} of the momentum distribution, and is distinct from the predicted thresholdless superradiance \cite{PhysRevLett.112.143002,PhysRevLett.112.143003, Chen_2014,Sandner:2015aa,Mivehvar_2017} induced by Fermi surface nesting, which relies on the {\it shape} of the Fermi surface. It is thus more robust to experimental imperfections such as finite temperature or inhomogeneities. Upon further increasing the density, recoil processes become Pauli-excluded and photon scattering is suppressed (Fig. \ref{fig:fig1}c), leading for example to sub-linear scaling of the threshold with atom number \cite{doi:10.1126/science.abd4385}. This behavior at zero-temperature is in sharp contrast with the case of weakly-interacting Bose-Einstein condensates, which displays no density dependence.

Our setup builds upon a cavity microscope described in \cite{Orsi_2024} and consists of a Fabry-Perot cavity and a confocal lens pair sharing a common optical axis. 
As shown in Fig. \ref{fig:fig1}d, the system comprises a pair of identical mirrors forming a near-concentric cavity, with high-reflectivity at \SI{671}{\nano\meter}, the main dipole-allowed transition of $^6$Li, yielding the cavity-QED parameters $(g/2\pi,\kappa/2\pi)=(2,1.1)$ \SI{}{\MHz} and a single-atom cooperativity of $\eta = 4g^2/\kappa\Gamma = 2.52$. The mirrors are also highly reflective at \SI{1342}{\nano\meter}, providing an intracavity optical dipole trap \cite{supp}. The mirror coatings feature a high-transmission window around \SI{765}{\nano\meter}, allowing tight optical-tweezer trapping of the atoms in the center of the cavity mode using the full available numerical aperture of $0.34$.

The sample preparation follows standard laser cooling and evaporative cooling in a cavity-assisted dipole trap \cite{supp} at the magnetic field of \SI{316}{\Gauss}, yielding a cold mixture of the first and third hyperfine levels (labeled $\ket{\downarrow}$ and $\ket{\uparrow}$). Figure \ref{fig:fig1}e shows a typical absorption image of the cloud produced after this step, with a temperature of $\mathrm{T}/\mathrm{T}_{\mathrm{F}}=1.1$ and a total atom number $\mathrm{N}=3\times10^4$. 
Performing the same evaporation procedure in the dipole trap in the presence of a tweezer trap at \SI{765}{\nano\meter} tightly focused at the center of the cavity yields about $4000$ atoms in the tweezer on top of an extended ultracold gas, as can be seen in Fig. \ref{fig:fig1}f. The tweezer has a trap depth of \SI{10}{\micro\kelvin}, a transverse trap frequency of \SI{40}{\kilo\hertz} and a beam waists of about \SI{0.95}{\micro\meter} estimated from the trap frequency ratio. Thermal equilibrium in the combined trap yields an expected $100$-fold decrease in $\mathrm{T}/\mathrm{T}_{\mathrm{F}}$ at the center of the tweezer \cite{PhysRevLett.81.2194,Serwane:2011aa}. We then turn off the dipole trap (Figure \ref{fig:fig1}g) and ramp down the tweezer-trap power to a variable set point to achieve the desired atom number. This procedure allows us to prepare atomic clouds with $40<\mathrm{N}<2100$ in a balanced mixture of $\ket{\uparrow}, \ket{\downarrow}$. 

\begin{figure}[t]
\includegraphics{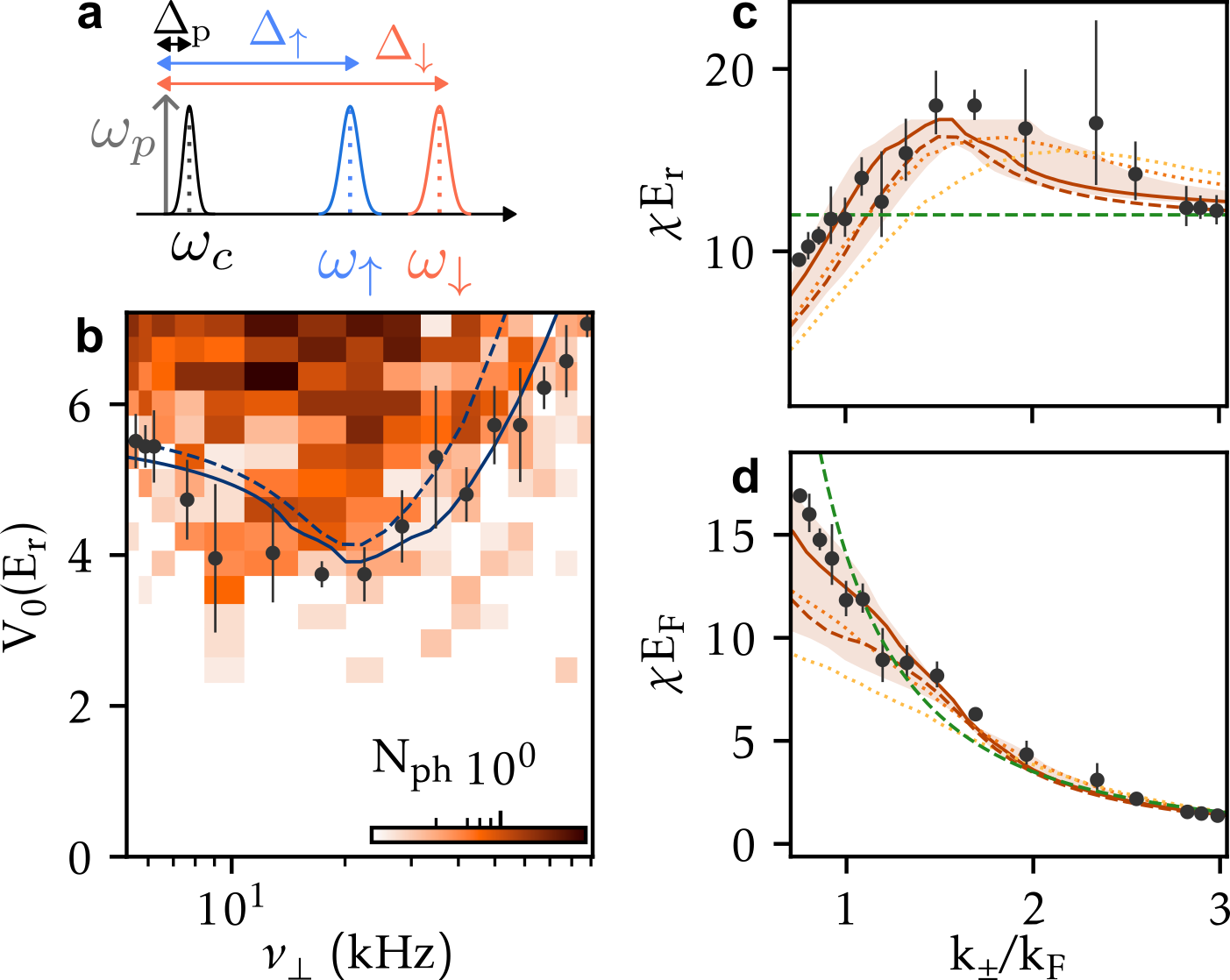}
    \caption{\label{fig:fig2} Superradiant phase transition varying atomic density. 
    (a) The two spin components $\ket{\downarrow},\ket{\uparrow}$ interact with the cavity with resonance frequency $\omega_{\mathrm{c}}$ and are driven by the pump field at frequency $\omega_{\mathrm{p}}$. (b)  Photon number $\mathrm{N_{ph}}$ detected within \SI{30}{\micro\second} following a quench of the pump laser power to $\mathrm{V}_0$, with varying trap frequency. 
    Unless otherwise stated, all uncertainties on atom number represent $68\%$ confidence interval on the mean, estimated using a resampling method \cite{supp}. Black points indicate the extracted thresholds. Error bars represent the standard deviation obtained by varying the critical photon number by $\pm30\%$ \cite{supp}. Solid and dashed lines show theoretical predictions using the local density approximation (\raisebox{0.5ex}{\protect\tikz{\protect\draw[black, line width=0.4mm] (0,.8ex)--++(.3,0);}}) and exact harmonic-oscillator eigenstates (\raisebox{0.5ex}{\protect\tikz{\protect\draw[black, densely dashed, line width=0.4mm] (0,.8ex)--++(.3,0);}}). Each datapoint is averaged $9$ times.
    (c) Normalized susceptibility extracted from the threshold as a function of the reduced wavevector $\mathrm{k}_{\pm}/\mathrm{k}_{\mathrm{F}}$, showing a local maximum near $\mathrm{k}_{\pm}/\mathrm{k}_{\mathrm{F}} \sim 1.7$. The experimental data are compared to theory with zero-temperature LDA (\raisebox{0.5ex}{\protect\tikz{\protect\draw[LDA, line width=0.4mm] (0,.8ex)--++(.3,0);}}) and exact eigenstates (\raisebox{0.5ex}{\protect\tikz{\protect\draw[LDA, densely dashed, line width=0.4mm] (0,.8ex)--++(.3,0);}}), as well as finite-temperature LDA theory for $\mathrm{T}/\mathrm{T}_{\mathrm{F}}=0.3$ (\raisebox{0.5ex}{\protect\tikz{\protect\draw[LDA03, densely dotted, line width=0.4mm] (0,.8ex)--++(.3,0);}}) and $0.5$ (\raisebox{0.5ex}{\protect\tikz{\protect\draw[LDA05, densely dotted, line width=0.4mm] (0,.8ex)--++(.3,0);}}) and with the predictions for a Bose-Einstein condensate (\raisebox{0.5ex}{\protect\tikz{\protect\draw[BEC, densely dashed, line width=0.4mm] (0,.8ex)--++(.3,0);}}). The shaded area represents the effect of atom-number uncertainties on the zero-temperature LDA.
    (d) Susceptibility normalized by Fermi energy, compared with the same susceptibility predictions of panel (c).}
\end{figure}

We investigate the cavity superradiant transition by sending a retro--reflected, transverse pump laser beam with wavevector $\mathbf{k}_{\mathrm{p}}$, polarized perpendicularly to the cavity axis. The beam intersects the cavity axis at a $90^{\circ}$ angle at the location of the atoms, creating a lattice potential with depth $\mathrm{V}_{\mathrm{0}}$. The atoms scatter light from the pump beam into the cavity field and back, yielding recoils at wavevectors $\pm \mathbf{k}_{\pm} = \pm \mathbf{k}_{\mathrm{c}} \pm\mathbf{k}_{\mathrm{p}} $, $\mathrm{k}_{\pm} = \vert \mathbf{k}_{\pm}\vert = \sqrt{2}\vert \mathbf{k_{c}}\vert$, where $\mathbf{k}_{\mathrm{c}}$ is the cavity wavevector, as sketched in Figure \ref{fig:fig1}d. The atoms undergo a superradiant phase transition when $\mathrm{V}_0$ exceeds a critical value $\mathrm{V}_{0c}$, corresponding to the emergence of a macroscopic density-wave in the gas at $\pm \mathbf{k}_{\pm}$ and the simultaneous buildup of a macroscopic field amplitude in the cavity.
The level configuration and laser detunings are presented in Fig. \ref{fig:fig2}a. The pump laser is red-detuned from the TEM$_{00}$ mode of the cavity resonance by  $ \omega_c - \omega_{\mathrm{p}}= \Delta_{\mathrm{p}} = 2\pi\times$\SI{1.2}{\mega\hertz}, chosen similar to the cavity linewidth to minimize $\mathrm{V}_{0c}$. There, the cavity timescales $\sim 1/\kappa \sim 1/\Delta_{\mathrm{p}}$ are much faster than the atomic scales. For atoms in state $\ket{\uparrow}$ (respectively $\ket{\downarrow}$), the pump has a detuning $\Delta_{\uparrow(\downarrow)} = \omega_{\uparrow(\downarrow)} - \omega_{\mathrm{p}} =$ $2\pi\,\times\,$\SI{300(465)}{\mega\hertz} with respect to the $\sigma_-$ transition to $\ket{2\mathrm{P}_{3/2},\mathrm{m_J}=-3/2,\mathrm{m_I}=-1 (+1)}$. The detunings were chosen positive to operate in the simplest configuration allowing for cavity superradiance \cite{mivehvar:2021aa}, with a magnitude allowing to reach $\mathrm{V}_{0c}$ with the available laser power. We denote as $\mathrm{V}_0$ the potential depth felt by atoms in state $\ket{\uparrow}$, which is twice larger than for atoms in state $\ket{\downarrow}$ due to the relative detunings. Both spins couple to the same cavity photons, yielding a single superradiant transition for the whole gas, with Fermi statistics affecting the overall threshold due to the phase space restrictions of each individual component.

Following the preparation of a mesoscopic gas, the pump laser is abruptly turned on at a strength $\mathrm{V}_0$ and the total number of photons $\mathrm{
N_{ph}}$ leaking through one of the cavity mirrors within \SI{30}{\micro\second} is recorded. With our choice of detunings, we estimate less than one spontaneous emission event within this timescale, allowing for a purely dispersive description. When $\mathrm{V}_0$ exceeds $\mathrm{V}_{0c}$, the cavity field builds up yielding finite stream of photons, indicating the emergence of the superradiant instability. This protocol measures $\mathrm{V}_{0c}$ with similar accuracy than the widely used adiabatic pump laser ramp (see \cite{PhysRevX.15.021089,marijanovic:2024aa} and \cite{supp} for details). To observe the effect of Fermi pressure on the transition, we repeat these measurements varying the tweezer trap depth over about two-orders of magnitude after sample preparation, hence the trap frequencies over one order of magnitude, keeping the atom number fixed at $\mathrm{N}=48\pm36$ (see \cite{supp} for details on atom numbers). This compression varies the Fermi wavevector $\mathrm{k_F}$ continuously between $0.33\,\mathrm{k_\pm}$ and $1.33\,\mathrm{k_{\pm}}$, with $\mathrm{k_F}=\sqrt{2m\mathrm{E_F}}/\hbar$, $\mathrm{E_F} = (3\mathrm{N})^{1/3}\hbar\bar\omega$, and $\bar\omega$ the geometric mean of the frequencies of the tweezer trap. The results are shown in Figure \ref{fig:fig2}b. The raw photon counts are presented in logarithmic scale, together with the transition threshold inferred from the photon counts \cite{supp}. The superradiant threshold first decreases as the trap frequency is increased, reaching a broad minimum before increasing again for the tightest confinements, manifesting the transition between free-space, Fermi-pressure-assisted and Pauli-blocked superradiance. 

\begin{figure*}[t]
\includegraphics{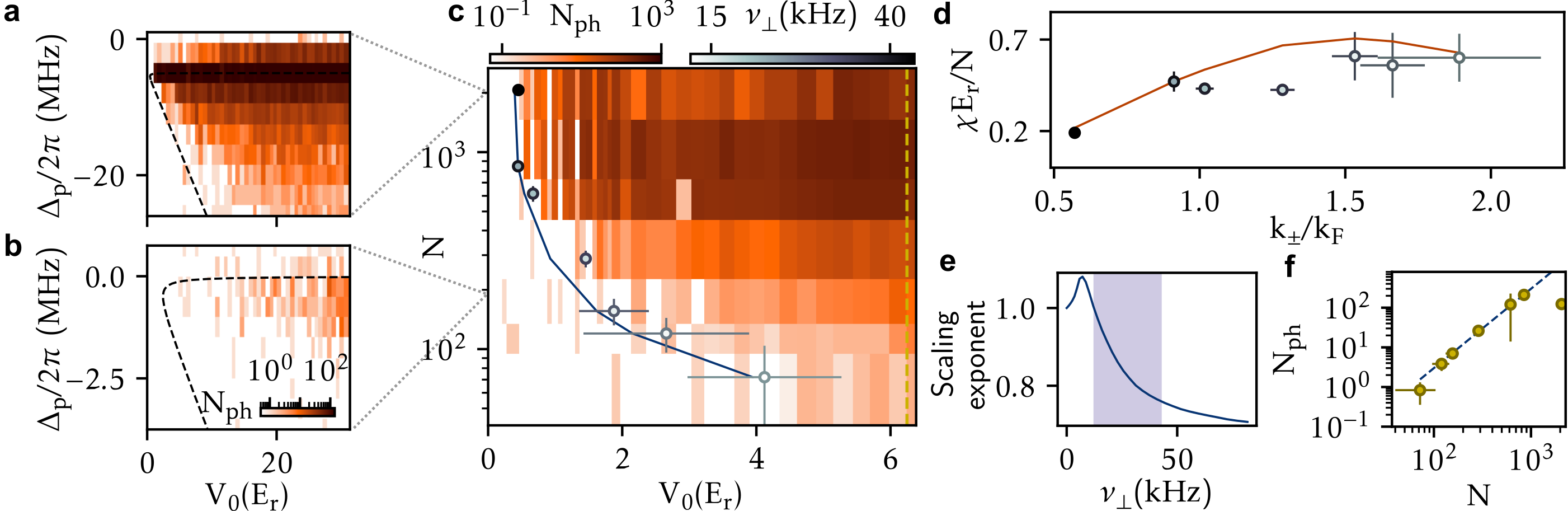}
\caption{\label{fig:fig3} Superradiant phase transition varying atom number. (a-b) Photon number recorded within \SI{30}{\micro\second} following a quench as a function of both $\mathrm{V_0}$ and $\Delta_{\mathrm{p}}$ plane for (a) $\mathrm{N}=1120\pm 120$, (b) $\mathrm{N} = 108\pm 60$. The dashed lines (\raisebox{0.5ex}{\protect\tikz{\protect\draw[fig4a, densely dashed, line width=0.4mm] (0,.8ex)--++(.3,0);}}) show the theoretical predictions based on Eq.\ref{eq:threshold} with LDA. Measurements are taken at $\Delta_{\uparrow}/2\pi=$\SI{414}{\mega\hertz}. Each data is averaged $2$ times. Both measurements are taken at the same tweezer trap depth of \SI{11}{\micro\kelvin}, resulting in $\mathrm{k_{\pm}/k_F}=(0.63,0.9)$. (c) Photon number with varying atom number within $70\leq\mathrm{N}<2100$ for $\Delta_{\uparrow}/2\pi=$\SI{400}{\mega\hertz}, $\tilde\Delta_{\mathrm{p}}/2\pi=$\SI{1.3}{\mega\hertz} and trap frequency. The grey scale represents the trap frequency used for the corresponding data point (scale on top of the figure). We overlap the measurement with the theory predictions (\raisebox{0.5ex}{\protect\tikz{\protect\draw[black, line width=0.4mm] (0,.8ex)--++(.3,0);}}) based on LDA simulation at zero temperature. The data is averaged between 2 and 10 times, depending on the signal-to-noise ratio. (d) Normalized susceptibility per atom extracted from the thresholds, compared with LDA predictions (\raisebox{0.5ex}{\protect\tikz{\protect\draw[LDA, line width=0.4mm] (0,.8ex)--++(.3,0);}}). (e) Scaling exponent $\alpha$ of the superradiant threshold $\mathrm{V_{0c}\propto N^{\alpha}}$, as a function of the perpendicular trap frequency, calculated using the local density approximation. The shaded region corresponds to the parameter range explored in panel (c). 
(f) Number of collected photons as a function of atom number for $\mathrm{V_0} = 6.2\,\mathrm{E_r}$ (\raisebox{0.5ex}{\protect\tikz{\protect\draw[photon_cut, densely dashed, line width=0.4mm] (0,.8ex)--++(.3,0);}} vertical cut in panel (c)), for which the threshold is exceeded for all atom numbers. Error bars on the photon number represent the standard deviation of the mean, calculated from different experimental repetitions. The dashed line is a guide to the eye showing the expected superradiant scaling $\mathrm{N_{ph}}\propto \mathrm{N^2}$.} 
\end{figure*} 

We now confirm this interpretation by quantitatively comparing our observations to theory. To this end, we model the atoms as a harmonically trapped Fermi gas at finite temperature, dispersively coupled to the single-mode cavity in the presence of an external pump. We perform a linear stability analysis developed for quench protocols in \cite{PhysRevX.15.021089,marijanovic:2024aa,supp}, leading to the following equation for the critical pump potential:
\begin{equation}\label{eq:threshold}
\mathrm{V}_{0c}= \frac{\tilde{\Delta}_{\mathrm{p}}^2 + (\frac{\kappa}{2})^2}{\tilde{\Delta}_{\mathrm{p}}} \frac{1}{2\chi U_0 (1+\zeta^2)}
\end{equation}
where $\tilde{\Delta}_{\mathrm{p}}$ is the detuning of the pump with respect to the cavity resonance, corrected from the bare detuning $\Delta_{\mathrm{p}}$ by the atoms-induced dispersive shift, $U_0=g^2/\Delta_{\uparrow}$ is the dipole potential created by one photon in the cavity and $\kappa$ is the cavity intensity decay rate. The factor $(1+\zeta^2)$ accounts for the unequal light-matter coupling of the two different spin components (see Eq. S18 in \cite{supp} for the definition of $\zeta$), which independently experience Fermi-pressure-assisted or Pauli-blocked scattering (Fig. \ref{fig:fig1}b-c) due to the fact that they do not interact. $\mathrm{V}_{0c}$ is the value of the pump potential $\mathrm{V_0}$ that triggers the phase transition. It is expressed in terms of $\chi$, the susceptibility measuring the linear response of the atomic density to the optical excitation at the wavevectors $\pm\mathbf{k}_\pm$. In our experimental protocol, where we quench the strength of the pump, this reduces to the zero-frequency density–density response of the Fermi gas, as captured by the Lindhard function \cite{osti_4405425}, measuring the phase space available for low-energy density excitations at a given momentum. Using our experimental parameters and performing trap averaging with the local-density approximation (LDA) yields a theoretical prediction for the zero-temperature threshold, shown as the solid line in Fig. \ref{fig:fig2}b-d. Here the pump strength $\mathrm{V_0}$ is calibrated using the weakest tweezer trap data, where the threshold is expected to be independent of quantum statistics or density. We find that it deviates by $40\%$ from a direct estimate based on the beam profile, laser power and atomic polarizability. For these weak trap data, anharmonic effects in the tweezer which may lead to a systematic overestimate of the Fermi wavevector do not influence the physics.

The results are in good agreement with our experimental observations. In particular, the occurrence and location of the broad minimum in the threshold is accurately reproduced. We can use the threshold observed in our data to infer the susceptibility as a function of the reduced parameter $\mathrm{k_\pm/k_F}$, as shown in Fig. \ref{fig:fig2}c, with the susceptibility normalized by the recoil energy $\mathrm{E_r} = \hbar^2\mathrm{k_{\pm}}^2/2m$. There we also present LDA results at finite temperature, where the maximum lowers and shifts to higher wavevectors due to the redistribution of momentum states occupation. The best agreement with our data is found for the zero temperature curve, providing an indirect confirmation of the low temperature of our mesoscopic Fermi gas. In the high-density regime (low $\mathrm{k_\pm/k_F}$), the suppression is due to superradiance competing with Fermi pressure. In fact, at higher temperature, an increase of the threshold persists when the recoil energy is much smaller than temperature \cite{supp,piazza:2013aa}. This behavior is in sharp contrast with the prediction for a Bose-Einstein condensate, where the susceptibility of the gas is given by the inverse of the recoil energy regardless of its density.

For our low atom number, we can also go beyond the LDA and describe the atoms using discrete energies and exact eigenfunctions to estimate response functions \cite{supp}. The predictions are shown with the dashed line in Fig. \ref{fig:fig2}b-d, deviating from the LDA for the tightest confinements, likely due to the trap frequency approaching the recoil. Lastly, normalizing the measured susceptibility by the Fermi energy provides a measurement of the trap-averaged Lindhard function as a function of the dimensionless momentum, showing a monotonic response at all momenta and the absence of Fermi surface nesting, as can be seen in Fig. \ref{fig:fig2}d. 

The mesoscopic atom number regime allows for the direct manifestation of the Fermi-pressure effects, but the physics is generic and persists for atom numbers in the macroscopic regime explored in previous experiments \cite{doi:10.1126/science.abd4385, Helson_2023}. We now leverage our mesoscopic preparation procedure to track superradiance over a wide range of atom number. Figure \ref{fig:fig3}a and b show two examples of phase diagrams collected in the $\Delta_{\mathrm{p}}$-$\mathrm{V_0}$ plane for $1120\pm120$ and $108\pm60$ atoms respectively. In both cases we observe the expected corner-shaped phase diagram, with a signal strongly reduced for the lowest atom numbers.

To systematically explore atom-number dependence, we fix $\tilde\Delta_{\mathrm{p}}=2\pi\times$\SI{1.3}{\mega\hertz}, corrected for the atom-number dependent dispersive shift, and acquire photon traces with varying atom numbers between $\mathrm{N}=70$ and $2100$ and trap frequencies, to span a wide range of $\mathrm{k_{\pm}/k_F}$. This choice of detuning minimizes the threshold and maximizes the number of photons. To further eliminate spurious atom-number dependent biases, the magnetic field is ramped to \SI{566}{\Gauss} reducing the atom-atom scattering length to zero. The results are shown in Fig. \ref{fig:fig3}c, together with the threshold inferred from the increase of photon number in the corresponding traces. The thresholds observed over the whole range of parameters are in reasonable agreement with the zero-temperature LDA theory. Similar to the density-dependence investigations in Fig. \ref{fig:fig2}, the variation of atom number also tracks the susceptibility as a function of $\mathrm{k_\pm/k_F}$. The results, normalized by atom number, are presented in Fig. \ref{fig:fig3}d, together with the theoretical predictions. 

Interestingly, our general theory predicts that the threshold does not follow a simple power-law with atom-number. However, a power-law fit over a limited parameter range provides insights into the effects of Fermi pressure, as used in \cite{doi:10.1126/science.abd4385}. Fig. \ref{fig:fig3}e presents the predictions for the scaling exponent as a function of trap frequencies for our geometry and the parameters of Fig.\ref{fig:fig2}. The regime of tight traps displays a sublinear scaling, with an exponent smoothly decreasing with density, in qualitative agreement with \cite{doi:10.1126/science.abd4385}, where Fermionic effects were favored by the presence of a deep pump-induced optical lattice. In addition, our model predicts a smooth increase of the scaling exponent with decreasing density, with a superlinear behavior in the Fermi-pressure-assisted regime, showing that a sublinear scaling is a fingerprint but not a necessary consequence of Fermi statistics.

Last, a distinctive feature of the superradiant phase is the super-linear scaling of the light intensity with atom number, due to the constructive interference of the light scattered by each of the atoms \cite{domokos:2002aa,yan:2023aa}. Figure \ref{fig:fig3}f shows the photon number observed in the superradiant phase as a function of atom number, for a fixed value of the pump strength $\mathrm{V_0}$, matching well the power law $\mathrm{N_{ph}}\propto \mathrm{N^2}$ expected for classical particles. This behavior persists as the integration time is varied \cite{supp}. Correction to the superradiant scaling of the photon number with $\mathrm{k_{\pm}/k_F}$, resulting from Pauli-blocking effects, are expected to be small in our parameter regime \cite{supp}. 

In conclusion, we have explored cavity-induced superradiance in a mesoscopic Fermi gas, charting its interplay with Fermi statistics. Our results reveal a non-monotonic threshold when compressing the gas at fixed atom number, due to the Fermi-pressure-induced occupation of finite momentum states at zero temperature. Interestingly, our findings are reminiscent of the interplay between temperature-induced momentum occupation and recoil processes predicted to yield a temperature-enhanced superradiance in Bose gases \cite{piazza:2013aa}, a feature so far never observed experimentally.  

In addition to allowing for the observation of the Fermi-pressure assisted superradiant transition, operating in the mesoscopic quantum gas regime has technical implications for future quantum simulations. First, as we show in End Matter, the mesoscopic atom number allows to operate in a regime where the collective Rabi frequency is smaller than the energy splitting between the spin states. As a result, we can operate in the strong dispersive regime with opposite coupling for the two spin components, endowing the ordered phase with an Ising-type spin-density wave character, different from the transverse spin waves produced so far using Raman coupling in Bose-Einstein condensates \cite{Zhiqiang:2017aa,Kroeze_2018,Landini:2018aa}. 
Second, for any cavity QED experiment in the optical regime, producing an entangled state of $\mathrm{N}$ atoms using cavity-mediated interactions requires a cooperativity $O(\mathrm{N}^2)$ down to $O(\mathrm{N})$ for heralded protocols \cite{Sorensen:2003aa}. Our mesoscopic Fermi gas opens the possibility to reach a break-even point where system-wide entanglement mediated by light could emerge in cavities with realistic cooperativities from few tens to hundreds \cite{jandura:2024aa,grinkemeyer:2025aa, bolognini:2025aa}. Apart from the creation of entanglement, the mesoscopic regime will make it possible to study genuinely quantum effects beyond mean-field in self-organization \cite{muller:2025aa}.
In addition, mesoscopic samples may allow for increasing the number of available cavity modes \cite{Vaidya:2018aa,kroeze:2025aa} beyond the number of particles, a prerequisite for the realisation of some quantum chaotic models  \cite{uhrich2023cavityquantumelectrodynamicsimplementation, baumgartner2024quantumsimulationsachdevyekitaevmodel, solis2025singleparticlemanybodychaosyukawasyk,baumgartner2025quantumsimulationusingtrotterized}. 
Lastly, our platform can also be scaled by projecting several tweezer traps, loaded from a single reservoir in order to investigate tunnel-coupled gases \cite{Murmann:2015aa,spar:2022aa,Jain:2025iyq} in the presence of strong light-matter interactions, generalizing to quantum matter the pioneering work performed with thermal atoms \cite{Ho_2025}.

The data sets generated and analyzed in the current study are available via Zenodo \cite{orsi_2026_18935490}.

{\it Acknowledgments --- } We thank Alexander Baumgärtner for discussions, Michał Zdziennicki, Tabea Bühler, Francesco Piazza, Michele Pini and Tobias Donner for careful reading of the manuscript and discussions. We acknowledge funding from the Swiss State Secretariat for Education, Research and Innovation (Grants No. MB22.00063 and 20QU-1\_215924) and the Swiss National Science Foundation (Grant No. 217124).

\providecommand{\noopsort}[1]{}\providecommand{\singleletter}[1]{#1}%

\begin{widetext}
\begin{center}
\section*{End Matter}
\end{center}
\end{widetext}

\begin{figure}[t]
\includegraphics{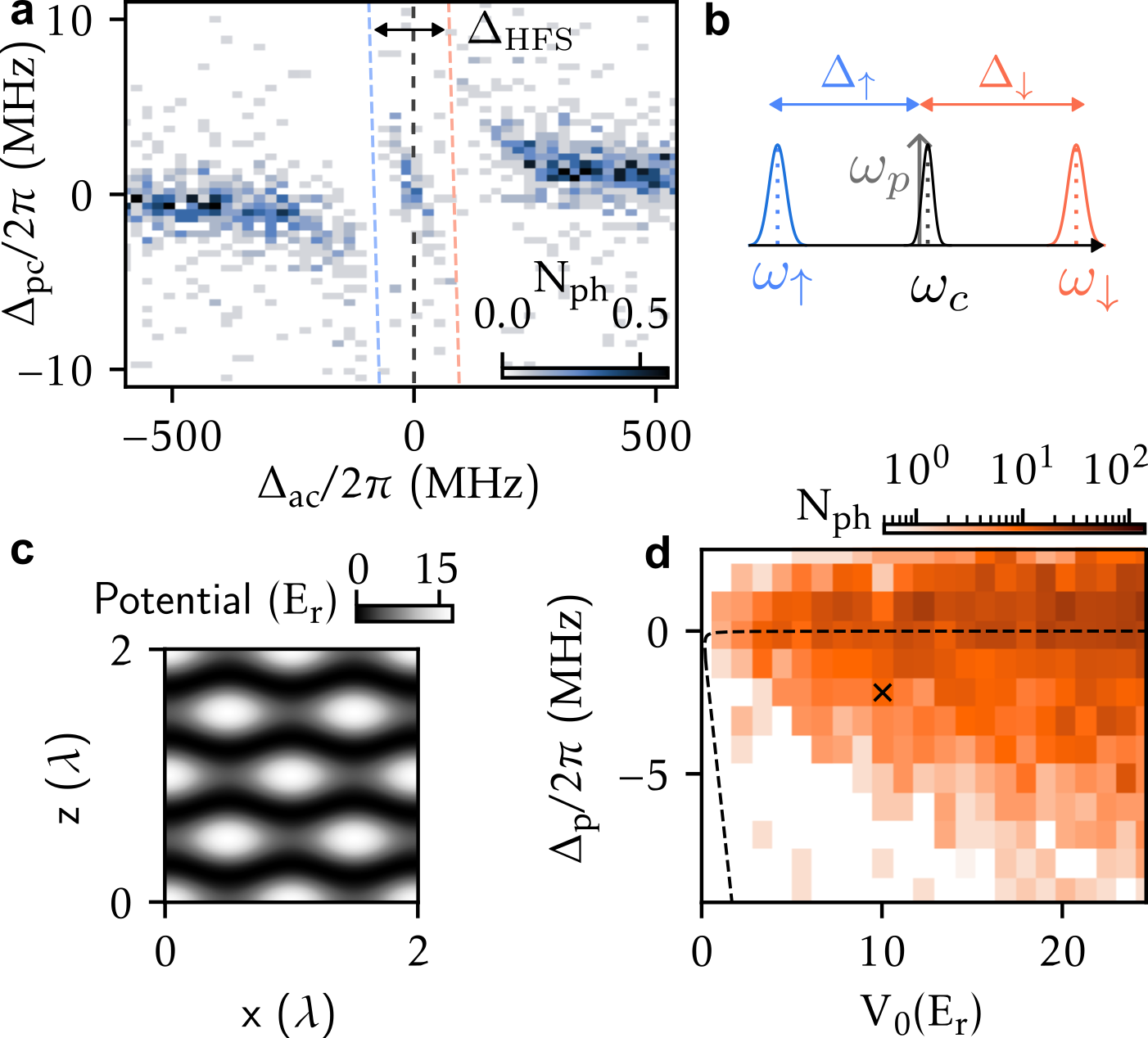}
\caption{\label{fig:fig4} Magnetic superradiance (a) Cavity transmission spectrum showing two avoided crossings corresponding to the cavity being on resonance with states $\ket{\uparrow}$ (\raisebox{0.5ex}{\protect\tikz{\protect\draw[statedown, densely dashed, line width=0.4mm] (0,.8ex)--++(.3,0);}}) and $\ket{\downarrow}$ (\raisebox{0.5ex}{\protect\tikz{\protect\draw[stateup, densely dashed, line width=0.4mm] (0,.8ex)--++(.3,0);}}) respectively. $\Delta_{\mathrm{pc}}$ refers to the detuning between the cavity probe and the cavity resonance and $\Delta_{\mathrm{ac}}$ is the detuning between the atomic resonances and the cavity resonance, and is set to zero when the cavity resonance is in-between the two atomic transitions. Each vertical line is averaged 15 times. To study magnetic superradiance, we fix the cavity resonance in between the two states (\raisebox{0.5ex}{\protect\tikz{\protect\draw[fig4a, densely dashed, line width=0.4mm] (0,.8ex)--++(.3,0);}}) as shown in (b). (c) Sketch of the potential produced by the combined pump and cavity fields in the superradiant phase, for a pump-cavity detuning $\Delta_{\mathrm{p}}/2\pi=$\SI{2}{\MHz} and a pump potential $\mathrm{V_0}=10\,\mathrm{E_r}$, where the potential in the cavity direction is extracted from the photon number of panel (d) (black cross). The potential is repulsive for $\ket{\uparrow}$ and attractive for $\ket{\downarrow}$. (d) Photon numbers detected as a function of $\Delta_{\mathrm{p}}$ and $\mathrm{V_0}$ showing superradiance in the magnetic regime for $\mathrm{N}= 520\pm 40$ atoms and the LDA prediction for the threshold (\raisebox{0.5ex}{\protect\tikz{\protect\draw[fig4a, densely dashed, line width=0.4mm] (0,.8ex)--++(.3,0);}}). Each data point is averaged between 2 and 4 times. }
\end{figure}

The mesoscopic character of our system allows us to study the cavity superradiant phase transition in a different regime, where the cavity field couples to the longitudinal magnetization of the Fermi gas, i.e. where light exerts an opposite dipole potential on the two spin components. Here, we report our experimental observations of this regime. 

Our Fermi gas comprises an equal mixture of two hyperfine states, separated by a frequency $\Delta_\mathrm{HFS}/2\pi$=\SI{165}{\mega\hertz}. So far, we have operated in a regime where the cavity field is red-detuned from both states (see Fig. \ref{fig:fig2}a), such that the photons exert forces of the same sign on both spin components. Consequently, the resulting superradiant phase has a charge density wave character \cite{bühler2025microscopycavityinduceddensitywaveordering}. We now tune the cavity resonance frequency half-way between the two hyperfine levels, as shown in Fig. \ref{fig:fig4}a and b. There, the existence of a dispersive regime requires the collective Rabi splitting to be lower than half the energy splitting between states, forcing the total atom number to be small, $\mathrm{N}\ll (\Delta_\mathrm{HFS}/4g)^2=3400$. The opposite detunings between light and atomic resonances for the two spin components imply that the simple scalar light-shift will tend to localize spin $\ket{\uparrow}$ atoms at minima and spin $\ket{\downarrow}$ at maxima of the field intensity, which is illustrated in Figure \ref{fig:fig4}c. This situation has similarities with two-components Bose-Einstein condensates driven by Raman transitions \cite{Zhiqiang:2017aa,Kroeze_2018} or in the presence of large vector polarizability \cite{Landini:2018aa}, but differs in fundamental ways. Raman transitions lead to transverse components of the magnetization, as opposed to our purely longitudinal, Ising-type spin-density wave. Vector polarizability produces differential light-shifts between spin components, but in \cite{Landini:2018aa} in dominates only along one particular direction selected by the pump polarization, such that density ordering remains present along the other direction. Importantly, none of these mechanisms are available for lithium atoms, for which the hyperfine coupling is too weak in the excited state manifold. 

Figure \ref{fig:fig4}a presents the transmission spectrum of the cavity containing a total of $\mathrm{N}=520\pm40$ atoms, as its resonance is scanned through the two hyperfine states. The measurement is obtained by shining a cavity probe laser along the cavity axis for \SI{150}{\micro\second} with varying $\Delta_{\mathrm{pc}}$, for different atom-cavity and probe-cavity detunings $\Delta_{\mathrm{ac}}$. We observe a clear transmission window between the two hyperfine resonances with zero net dispersive shift, indicating operation in the dispersive regime with opposite scalar shifts.

In this configuration, we repeat the protocol described in the main text and observe the onset of the superradiant phase at a magnetic field of \SI{566}{\Gauss}. The results are presented in Figure \ref{fig:fig4}d. Generalizing the stability analysis \cite{marijanovic:2024aa} to opposite spin coupling yields a phase boundary determined by the spin susceptibility of the system at the relevant wavevectors \cite{supp}. In contrast with density-wave ordering at similar atom number, here we find strong deviations from the theory. 
It is known that the linear stability analysis does not capture well the phase boundary for light producing a repulsive potential in a Bose-Einstein condensate \cite{Zupancic:2019wr}, in which superradiance only exists within a narrow detuning range where the frequency difference between the pump and the dispersively-shifted cavity resonance is smaller than the dispersive shift of the cavity. However, the phase diagram in Fig. \ref{fig:fig4}d is qualitatively similar to that of red-detuned cavities (Fig. \ref{fig:fig3}a-b), calling for a finer theoretical understanding of this new regime. In the strongly interacting case, spin-density wave order would directly compete with pairing, an intriguing situation not envisioned so far.

\newpage

\pagebreak
\clearpage
\renewcommand{\theequation}{S\arabic{equation}}
\renewcommand{\thefigure}{S\arabic{figure}}
\setcounter{figure}{0}

\section*{Supplemental Material}
\setcounter{page}{1}
\renewcommand{\thepage}{S\arabic{page}}

\subsection{Experimental apparatus}\label{sec:experimental_apparatus}
The core of the experimental apparatus is the cavity-microscope device described in \cite{Sauerwein_2023, Orsi_2024}. The optical cavity has a finesse of \num{5265} at \SI{671}{\nano\meter}, near--resonant to the D\num{2} transition of $^6$Li, and a finesse of \num{19306} at \SI{1342}{\nano\meter}. The latter wavelength is used both to stabilize the cavity length and to generate a deep intracavity dipole trap. The cavity length is \SI{25.987}{\milli\meter}, corresponding to a free spectral range of $\nu_{\mathrm{FSR}}=$\SI{5.77}{\giga\hertz}, and it is operated approximately \SI{100}{\micro\meter} from concentric. At \SI{671}{\nano\meter}, this results in a cavity linewidth of $\kappa/2\pi = $\SI{1.1}{\mega\hertz} and a maximum single--atom single--photon coupling $g/2\pi = $\SI{2}{\mega\hertz} at \SI{671}{\nano\meter}, corresponding to a cooperativity $\eta = 4g^2/\kappa\Gamma$= \num{2.52}.

The lenses in the mirror--lens assemblies have a numerical aperture of \num{0.345}. The lenses are optimized to be aberration-free at \SI{780}{\nano\meter}. The wavelength of the tweezer beam is monitored with a wavemeter and set to \SI{765}{\nano\meter} where we found transmission through the cavity mirrors to be maximum. From parametric heating, we measure a transverse to longitudinal trap frequency ratio of \num{5.5}, from which we infer a tweezer waist of \SI{0.95}{\micro\meter}, close to the aberration-free gaussian beam waist $0.84 \lambda f/d = $ \SI{0.87}{\micro\meter}, calculated from the lens focal length $f=$\SI{19.06}{\milli\meter} and clear aperture of $d=$\SI{14}{\milli\meter}. From the measurement of the beam waist we extract a Rayleigh range of \SI{3.68}{\micro\meter}. 

We operate the experiment with a single laser at \SI{1342}{\nano\meter}, whose frequency is stabilized with a Pound--Drever--Hall (PDH) lock on the cavity resonance. The PDH light acts at the same time as the intra--cavity dipole trap. Additional noise reduction of the intracavity dipole trap is achieved by fast feedback of the PDH error signal onto an electro-optic modulator, following methods similar to \cite{Endo:18, Vezio:25}. The same modulator is also used to generate sidebands for canceling the intracavity lattice \cite{PhysRevA.74.063401,PhysRevA.94.061601, PhysRevA.101.043611, Roux_2021, shadmany2024cavityqedhighna, B_hler_2025}. During evaporative cooling, the intracavity field is reduced using an acousto-optic modulator, and we compensate the reduction of PDH signal using a separate modulator acting before the photodiode. Liquid-crystal attenuators placed in front of the photodiodes increase the available range of power for which we can lock the cavity.

The \SI{1342}{\nano\meter} laser is frequency doubled to \SI{671}{\nano\meter} using a second--harmonic--generation cavity (Toptica TA-SHG). This light is used for zeeman slowing, magneto-optical trapping, absorption imaging, cavity probing, and transverse pumping. Control of the atom–cavity detuning is achieved using a beat-note lock to another laser stabilized onto lithium using saturated absorption spectroscopy, with feedback onto the cavity length.

\subsection{Experimental sequence}\label{sec:preparation_atoms}
The experimental sequence begins by loading a magneto-optical trap of $^6$Li atoms for $1.5$ seconds near the center of the cavity, using light near resonant with the D\num{2} transition. The magnetic compensation coils are then adjusted to overlap the atomic cloud with the cavity mode center in the presence of both the intracavity dipole trap and the optical tweezer, with respective trap depths of about \num{1240} and \SI{10}{\micro\kelvin}. To maximize loading efficiency, the lattice structure formed by the intracavity dipole trap is canceled by phase-modulating the incident at a frequency equal to the free spectral range. The magneto-optical trap is subsequently compressed over \SI{16}{\milli\second} by simultaneously reducing laser powers, magnetic field gradients, and detunings. After switching off the cooling light and magnetic fields, approximately one million atoms remain trapped at the cavity center in a mixture of the two lowest hyperfine states $\ket{1}, \ket{2}=\ket{^2\mathrm{S}_{F = 1/2}, \mathrm{m_F} = \pm 1/2}$.\\
The magnetic field is ramped to \SI{316}{\Gauss}, and atoms are transferred from state $\ket{2}$ to state $\ket{3} = \ket{^2\mathrm{S}_{\mathrm{F} = 3/2}, \mathrm{m_F} = - 3/2}$ using a radiofrequency Landau–Zener sweep from $85$ to \SI{90}{\mega\Hz} over \SI{1}{\milli\second}, with an efficiency of $85$\%. At this field, the scattering length between states $\ket{1}$ and $ \ket{3}$ is $a = $ \num{-890.5}$a_0$ \cite{Z_rn_2013}. In the main text, we refer to the hyperfine states as $\ket{1}= \ket{\downarrow}$ and $\ket{3} = \ket{\uparrow}$. Evaporative cooling is then performed using four piecewise linear ramps of the intracavity trap power over a total duration of \SI{500}{\milli\second} resulting in approximately 30000 atoms at a temperature of \SI{400}{\nano\kelvin}. This corresponds to a $\mathrm{T/T_F}=$ \num{1.14} for atoms in the cavity dipole trap, which has a final trap depth of \SI{1.38}{\micro\kelvin}, and to \num{0.01} at the bottom of the tweezer trap. We measure an upper bound for the temperature of the atoms in the cavity dipole trap with standard time--of--flight technique, and we assume that the atoms are in thermal equilibrium. After switching-off the intracavity dipole trap, we obtain a degenerate Fermi gas of approximately \num{4000} atoms trapped in a tweezer at the center of the cavity mode, in a balanced spin mixture. For measurements in the non-interacting regime, the magnetic field is ramped to \SI{566}{\Gauss} over \SI{200}{\milli\second}, where the scattering length is near zero \cite{Z_rn_2013}, while for interacting measurements it is held constant. The tweezer power is adjusted to vary the atom number and subsequently increased to reach the desired Fermi wavevector, varying the transverse trap frequency between $5$ and \SI{90}{\kHz}. 
We find that the atom number remains unchanged for a bias magnetic field of \SI{316}{\Gauss} and \SI{566}{\Gauss}, indicating that thermalization plays a negligible role when the tweezer power is lowered and that the tweezer-trap ramp primarily induces controlled spilling of an already highly degenerate quantum gas.

To trigger the superradiant phase transition, the power of the transverse side pump is quenched from zero to its target value within \SI{4}{\micro\second}. Photons leaking from the cavity are detected using a single-photon counting module. The side pump is detuned from both the empty cavity resonance and the atomic resonance of state $\ket{\uparrow}$, and has a waist of \SI{89}{\micro\meter}, measured by direct imaging in the atomic plane. After all measurements are performed, all traps are released and the empty cavity resonance is measured using on-axis transmission spectroscopy.
\subsection{Data processing}
\subsubsection{Extraction of the threshold}\label{sec:threshold_extraction}
 
We determine the critical pump power by recording the total number of detected photons leaking from the cavity over the first \SI{30}{\micro\second} following the quench. We found that increasing this duration improves the signal to noise ratio at the expense of a visible decay in photon population, likely originating from atom losses in the superradiant state, while reducing it reduces the signal below the noise floor.

Because of the small atom number and limited photon collection efficiency, the total photon count recorded per run of the experiment is low and subject to significant noise, primarily photon shot noise and shot-to-shot atom number fluctuations. To mitigate these effects on the threshold determination, we use a Gaussian filter on the photon count data to average neighboring values of $\mathrm{V_0}$. We then determine the pump power at which the signal exceeds a critical photon number. The robustness of this procedure is verified by varying the critical photon number by $\pm30\%$ and tracking the resulting critical pump power. The error bars shown correspond to the mean and standard deviation of the critical values obtained over this range. If the standard deviation is zero, we instead use the bin width to represent the uncertainty. 

For measurements of Figure 3c where the atom and photon number varies by more than one order of magnitude, the threshold value is adjusted accordingly. We verify that the extracted critical pump strength remains robust against variations in both the threshold choice and the signal integration time.

\subsubsection{Uncertainty on atom numbers}\label{sec:atom_number}

Atom numbers are measured using absorption imaging after a \SI{60}{\micro\second} time of flight and with a pulse duration of \SI{10}{\micro\second}. The imaging is calibrated against the dispersive shift of the cavity resonance at high atom numbers. The dispersive shift measurement fails at low atom number where it becomes smaller than the cavity linewidth. Thus for the data shown in Figure 3c, atom numbers are averaged between absorption imaging and dispersive shift measurements. For Figures 2, 3a-b and 4 only absorption imaging is used.

For low atom numbers, we estimate the uncertainty in atom numbers using a bootstrapping procedure on the absorption images. We believe that the stated uncertainty for low atom numbers is dominated by detection noise rather than sample preparation, due to the relatively good stability of the superradiance data. The resulting standard deviation of the mean is compared to that obtained from single-image measurements in a higher atom number regime, where reliable atom counting from individual images is possible. This analysis shows excellent agreement between the mean atom numbers extracted using both methods, as well as good agreement in the estimated uncertainties. 
\subsubsection{Photon number scaling}\label{sec:photon_scaling}

\begin{figure}[t]
\includegraphics{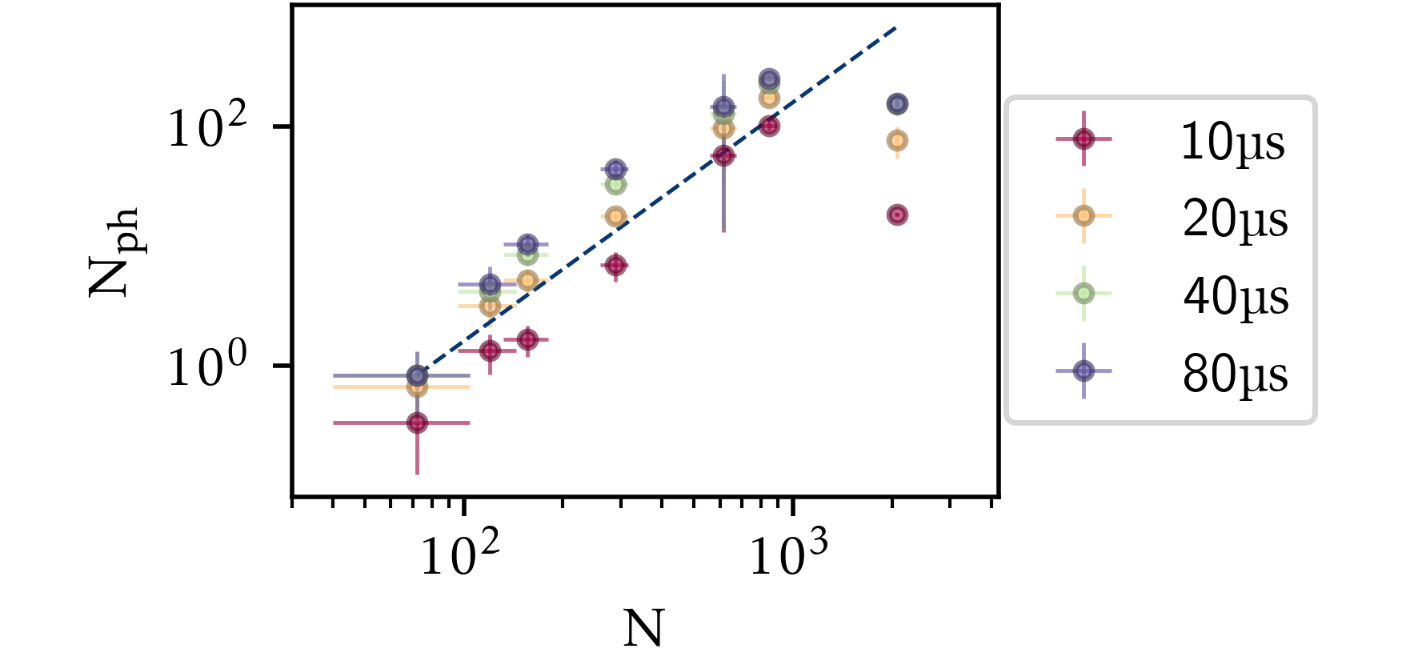}
\caption{\label{fig:M1} Photon number scaling from Figure 3c for different collection times after the quench. Increasing the collection time changes the offset of the signal but does not affect the scaling with atom number. Error bars on the photon number represent the standard deviation of the mean, calculated from different experimental repetitions.} 
\end{figure}

Figure \ref{fig:M1} shows the photon number collected by the single-photon counter after the onset of superradiance, plotted as a function of atom number for different collection times. Increasing the collection time shifts the absolute photon count but does not affect the scaling, confirming the observed atom-number dependence.

\subsection{Theoretical model} 

\subsubsection{Critical point} \label{sec:critical_point}

To describe the onset of cavity superradiance following a quench of the transverse pump, we apply the linear-response treatment developed in \cite{PhysRevX.15.021089, marijanovic:2024aa}. For the moment we treat the gas as spinless. We start from the cavity-QED Hamiltonian in the dispersive regime,\begin{align}\label{eq:hamiltonian1}
    \hat{\mathrm{H}}& = \tilde{\Delta}_{\mathrm{p
    }}\hat{a}^{\dagger}\hat{a} - \frac{g\Omega}{2\Delta}\hat{\theta} (\hat{a}+ \hat{a}^{\dagger}) 
\end{align}
where $\tilde{\Delta}_{\mathrm{p}}$ is the detuning of the pump from the dispersively shifted cavity resonance and $\hat{a}^{\dagger}, \hat{a}$ the raising and lowering operators for the cavity field. $g$ is the single-atom single-photon coupling strength of the atomic transition to the cavity field and $\Omega$ the Rabi frequency of the pump. $\Delta$ is the pump detuning from the atomic transition. The density-wave operator
\begin{align}\label{eq:def_theta}
\hat{\theta} = \int d\mathbf{r} \; \hat{n}(\mathbf{r}) \cos{(\mathbf{k_c \cdot r})}\cos{(\mathbf{k_p \cdot r})}
\end{align}
captures the coupling between the atomic density modulation and the interference of the pump with the cavity field. Here, $\hat{n}(\mathbf{r})$ is the fermionic density operator, defined as
\begin{equation}
\hat{n}(\mathbf{r}) = \hat \psi_g^{\dagger}(\mathbf{r}) \hat \psi_g(\mathbf{r}) = \sum_{ij} \phi_i^*(\mathbf{r}) \phi_j(\mathbf{r}) \hat c_i^{\dagger} \hat c_j,
\end{equation}
where $\hat \psi_g(\mathbf{r})$ is the field operator for atoms in the ground state, $\hat c_i$ ($\hat c_i^\dagger$) are the corresponding annihilation (creation) operators satisfying the fermionic anticommutation relations $\{\hat c_i , \hat c^{\dagger}_j\} = \delta_{ij}$ with $\delta_{ij}$ the Kronecker delta, and $\phi_i(\mathbf{r})$ are the single-particle eigenfunctions of the unperturbed system. This eigenbasis will be used in Section \ref{sec:simulation_susceptibility} to calculate the density-density response function.
 \\

Introducing the cavity field quadratures
\begin{equation}
    \left\{
    \begin{aligned} 
        & \hat{X } = \frac{1}{\sqrt{2}}(\hat{a}+ \hat{a}^{\dagger}) \\
     & \hat{P} = \frac{1}{\sqrt{2}i}(\hat{a}- \hat{a}^{\dagger})
     \end{aligned}
     \right.
\end{equation}
the Hamiltonian becomes 
\begin{equation}
    \hat{\mathrm{H}}=\frac{\tilde{\Delta}_{\mathrm{p}}}{2} (\hat{X}^2 + \hat{P}^2)- \frac{\sqrt{2}g\Omega}{2\Delta}\hat{\theta} \hat{X}
\end{equation}

The dynamics after the quench is described by the Heisenberg-Langevin equations for the cavity field, combined with linear response for the atomic degrees of freedom described by the Kubo formalism \cite{doi:10.1143/JPSJ.12.570}. This yields the coupled equations for the expectation values
\begin{equation}\label{eq:set_of_equations}
    \left\{
    \begin{aligned} 
        & \frac{d\langle \hat{X} \rangle}{dt}  =\tilde{\Delta}_{\mathrm{p}}\langle \hat{P} \rangle -\frac{\kappa}{2}\langle \hat{X} \rangle \\
     & \frac{d\langle \hat{P} \rangle}{dt} =  -\tilde{\Delta}_{\mathrm{p}}\langle \hat{X} \rangle-\frac{\kappa}{2}\langle \hat{P} \rangle+\frac{\sqrt{2}g\Omega}{2\Delta}\langle \hat{\theta}(t)\rangle   \\
     & \langle \hat{\theta}(t)\rangle =  \frac{\sqrt{2}g\Omega}{2\Delta} \int_0^\infty \chi(t- t') \langle \hat{X}(t')\rangle dt'
     \end{aligned}
     \right.
\end{equation}
where we used $\langle \hat{\theta} (t=0)\rangle=0$  in the absence of pumping. We define $\chi$ as the density-wave susceptibility of a single spin component
\begin{equation}\label{eq:def_susceptibility}
\chi(t- t') = -i \Upsilon(t-t')\langle [ \hat{\theta}(t),\hat{\theta}(t')]\rangle,
\end{equation}
with $\Upsilon(t)$ the Heaviside step function.

For the early-time evolution, the susceptibility is that of the trapped Fermi gas prior to the quench, in the absence of pump-induced optical lattice. The superradiant threshold is the critical $\Omega$ above which the cavity field $\langle \hat{X} \rangle$ grows exponentially, a manifestation of the instability of the $\langle \hat{X} \rangle=0$ solution. This is equivalent to the sign-change of the curvature of the Landau free-energy in the equilibrium picture. 

A detailed treatment of the quench dynamics can be found in \cite{marijanovic:2024aa}. In short, using the ansatz $\langle \hat{X} \rangle = X_0e^{-i z t}$, the threshold is determined from the emergence of a non-trivial solution to Eq. \ref{eq:set_of_equations} with $\mathrm{Im}(z)>0$, corresponding to the order growth rate. The critical pump strength is obtained from the condition $\mathrm{Im}(z)\longrightarrow0^+$. 

It yields the threshold condition in terms of the zero-frequency susceptibility:

\begin{align}\label{eq:threshold_single_species}
    \chi(\omega\longrightarrow0) &= \frac{\tilde{\Delta}_{\mathrm{p}}^2 +(\kappa/2)^2}{\tilde{\Delta}_{\mathrm{p}}} \frac{2 \Delta^2}{g^2\Omega^2}\\
    & = \frac{1}{2D_{0c}}.
\end{align}

 For simplicity, we denote this quantity as $\chi$. We have introduced an effective cavity-induced non-linearity 
\begin{equation}
    D_{0c} = \frac{\tilde{\Delta}_{\mathrm{p}}\mathrm{V}_{0c}U_0}{\tilde{\Delta}_{\mathrm{p}}^2 +(\kappa/2)^2},
\end{equation}

where $\mathrm{V}_{0c} = \Omega_c^2/4\Delta$ the critical pump-induced lattice depth and $U_0 = g^2/\Delta$ the dispersive shift per cavity photon.

In the experiment, we record the time integral of the photon number following the quench up to $T=$\SI{30}{\micro\second}, $\mathrm{N_{ph}} \propto \int_0^T \left| \langle \hat{X}(t) \rangle \right|^2 dt $. Below threshold, the initial condition $\langle \hat{X} \rangle=0$ ensures that $\mathrm{N_{ph}}=0$. The superradiant phase is then unambiguously identified from $\mathrm{N_{ph}}>0$. Strictly speaking, our finite integration time provides only an upper-bound to the threshold because close enough to the critical point, the growth rate may exceed the integration time. Setting $\mathrm{Im}(z) \sim \hbar/T$ and using the results of \cite{marijanovic:2024aa} for the BEC, corresponding to low-density regime of our experiment, we estimate our systematic bias on the threshold to be on the order of a few percent, way below our error bars.

\subsubsection{Two-component gas}
We now generalize the above treatment to a two-component Fermi gas with
potentially unequal light–matter coupling strengths. Each spin component
contributes with its own density-wave susceptibility $\chi$, while the cavity couples to
a collective density modulation involving both components. Introducing density-wave operators $\hat{\theta}_{i}$ for each spin component $i=\uparrow,\downarrow$, the Hamiltonian becomes
\begin{equation}
     \hat{\mathrm{H}} = \tilde{\Delta}_{\mathrm{p}}\hat{a}^{\dagger}\hat{a} - \frac{g_\uparrow\Omega_\uparrow}{2\Delta_\uparrow}\left( \hat{\theta}_\uparrow + \zeta\hat{\theta}_\downarrow \right) (\hat{a}+ \hat{a}^{\dagger}),
\end{equation}

expressed as a function of the dimensionless asymmetry parameter 
\begin{equation}\zeta= \frac{\Delta_\uparrow g_\downarrow\Omega_\downarrow}{\Delta_\downarrow g_\uparrow\Omega_\uparrow}.\end{equation}

We introduce the collective operator $\hat\Theta = \hat{\theta}_\uparrow + \zeta\hat{\theta}_\downarrow$, which couples linearly to the cavity field. The cavity therefore probes an effective collective susceptibility, which we
denote by $\chi_{\mathrm{eff}}$, defined as
\begin{align}\label{eq:suscept_eff}
\chi_{\mathrm{eff}}(t- t') &= -i \Upsilon(t-t')\langle [ \hat{\Theta}(t),\hat{\Theta}(t')]\rangle 
\end{align}

Since the two spin components are independent in our experiment and the gas is
balanced, both components have identical single-particle response functions $\chi$,
yielding
\begin{align}
\chi_{\mathrm{eff}} = (1+\zeta^2)\chi.
\end{align}
This defines the effective susceptibility $\chi_{\mathrm{eff}}$ that replaces $\chi$ in the single-component threshold condition Eq. \ref{eq:threshold_single_species}, yielding to the threshold equation 
\begin{align}\label{eq:threshold_double_species}
    \chi= \frac{1}{2D_{0c}(1+\zeta^2)}.
\end{align}

\subsubsection{Polarization effects}

The transverse pump is linearly polarized perpendicular to both the cavity axis and the magnetic field. In our parameter regime, the Zeeman splitting is non negligible compared to the detuning between the cavity resonance and the $\sigma_-$ transition. As a result, the contributions of the $\sigma_-$ and $\sigma_+$ polarizations to the light–matter coupling must be treated separately.\\

The Hamiltonian can then be written as
\begin{align}
    \hat{\mathrm{H}}   =&  \tilde{\Delta}_{\mathrm{p}}\hat{a}^{\dagger}\hat{a} -  (\frac{g_{\uparrow,+}\Omega_{\uparrow,+}}{2\Delta_{\uparrow,+}} + \frac{g_{\uparrow,-}\Omega_{\uparrow,-}}{2\Delta_{\uparrow,-}} ) \left( \hat{\theta}_\uparrow + \zeta\hat{\theta}_\downarrow \right) (\hat{
    a}^{\dagger}+\hat{
    a} ), 
\end{align}
with
\begin{equation}
    \zeta = \frac{\frac{g_{\downarrow,+}\Omega_{\downarrow,+}}{2\Delta_{\downarrow,+}} + \frac{g_{\downarrow,-}\Omega_{\downarrow,-}}{2\Delta_{\downarrow,-}}}{\frac{g_{\uparrow,+}\Omega_{\uparrow,+}}{2\Delta_{\uparrow,+}} + \frac{g_{\uparrow,-}\Omega_{\uparrow,-}}{2\Delta_{\uparrow,-}}}.
\end{equation}
Here, the subscripts $\uparrow,\downarrow$ refer to the spin species, while $+,-$ to the polarization components of the light. 
Solving the linearized equations of motion yields the threshold condition

\begin{align}\label{eq:threshold_polarizations}
\chi = \frac{\tilde{\Delta}_{\mathrm{p}}^2 + (\kappa/2)^2}{\tilde{\Delta}_{\mathrm{p}} (\frac{g_{\uparrow,+}\Omega_{\uparrow,+}}{2\Delta_{\uparrow,+}} + \frac{g_{\uparrow,-}\Omega_{\uparrow,-}}{2\Delta_{\uparrow,-}} )^2 (1 + \zeta^2 )}.
\end{align}
This equation provides the quantitative relation between the experimentally
measured threshold and the single-component susceptibility $\chi$ calculated from first principles, with all polarization and coupling asymmetries taken into account. In practice, the $\sigma_+$ transition is detuned by \num{1} and \SI{2}{\GHz} more than the $\sigma_-$ transition, depending on the magnetic field, such that its contribution is below \num{5}\% of the total light-matter coupling.

\begin{figure}[t]
\includegraphics{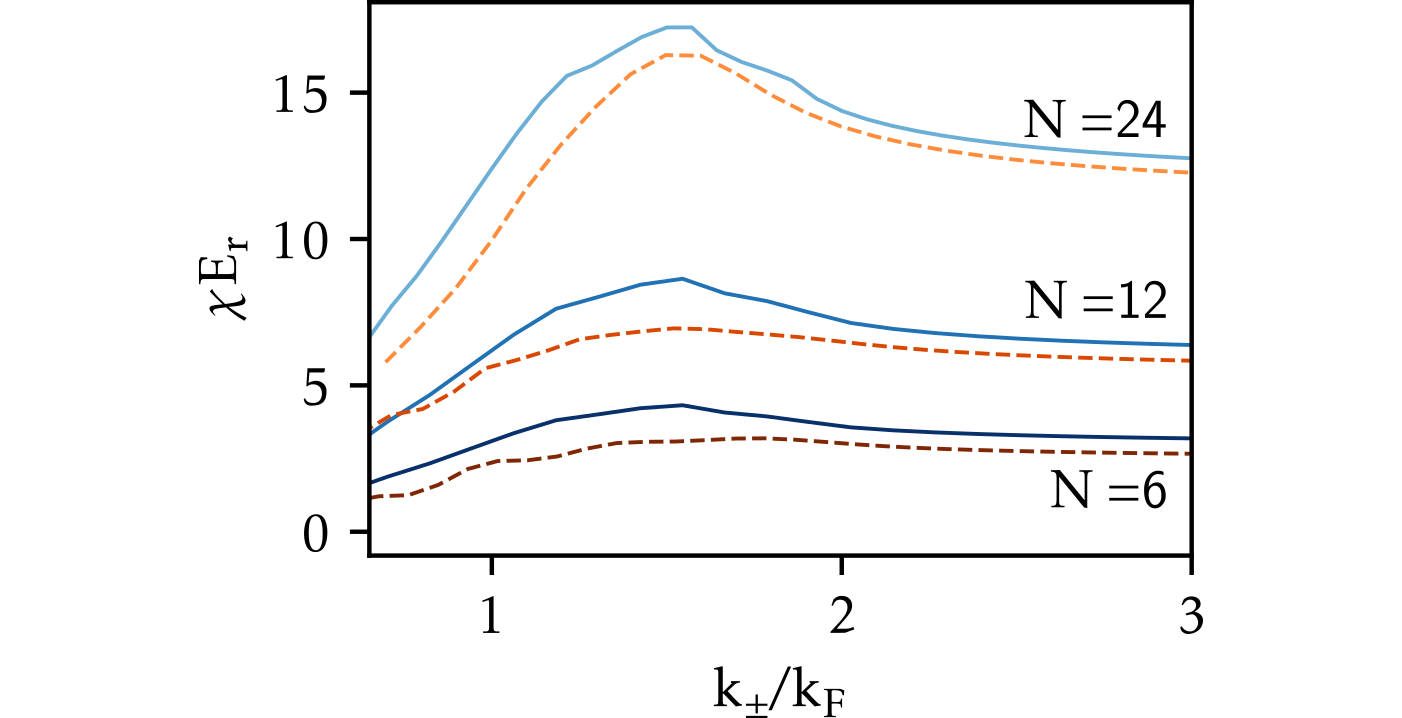}
\caption{\label{fig:epsart} Susceptibility for a single-specie Fermi gas calculated using the local density approximation (solid lines) and the exact eigenstates of the harmonic trap (dashed lines) for different atom numbers.} 
\end{figure}

\subsection{Simulations of the density-density response function}\label{sec:simulation_susceptibility}
To predict the phase boundary between the normal and superradiant phases, we calculate the zero-frequency density-density response function defined in Equations \ref{eq:set_of_equations}-\ref{eq:def_susceptibility}.  In this work, we evaluate $\chi$ using (I) plane-wave eigenstates in the local density approximation (LDA) and (II) the exact harmonic oscillator eigenstates.

\subsubsection{Local density approximation}
The susceptibility defined in Eq. \ref{eq:def_susceptibility} describes the response of the operator $\hat\theta$ at time $t$ to a perturbation at time $0$ coupled to the operator $\hat\theta$ in the context of linear response theory \cite{doi:10.1143/JPSJ.12.570}. To calculate the commutator, we can express the density wave operator from Eq. \ref{eq:def_theta} as
\begin{equation}
    \hat \theta = \frac{1}{4}\sum_{\mathbf{k} = \pm \mathbf{k}_{\pm}} \int d\mathbf{r}\hat{n}(\mathbf{r})e^{i \mathbf{kr}}=  \frac{1}{4}\sum_{\mathbf{k} = \pm \mathbf{k}_{\pm}} \hat{n}_{\mathbf{k}},
\end{equation}
where $\hat{n}_{\mathbf{k}}$ is the Fourier transform of the density operator $\hat{n}_{\mathbf{k}} = \int d\mathbf{r}\hat{n}(\mathbf{r})e^{i \mathbf{kr}}$. 
The commutator between $\hat\theta$ operators can be expressed in the more familiar commutator between density fluctuations writing 
\begin{align}\label{eq:commutator_theta}
    [\hat \theta (t), \hat \theta (0)] &= \frac{1}{16}\sum_{\mathbf{k} = \pm \mathbf{k}_{\pm}} [\hat{n}_{\mathbf{k}}(t),\hat{n}_{\mathbf{k}}(0)],
\end{align}
which implies an averaging of the response over the different wavevectors of the perturbation. To calculate the response in the local density approximation, we divide the 3D harmonic trap into small homogeneous volumes $\mathcal{V}$. The trap is treated as locally flat and the eigenstates in each volume are approximated by plane waves. In each volume, the response of the gas is isotropic, meaning that Eq. \ref{eq:commutator_theta} becomes 
\begin{align}
    [\hat \theta (t), \hat \theta (0)] =   \frac{1}{4}[\hat{n}_{\mathrm{k}_{\pm}}(t),\hat{n}_{\mathrm{k}_{\pm}}(0)],
\end{align}
where we have defined $\mathrm{k}_{\pm} = \vert \mathbf{\pm k_{\pm}}\vert$. For plane waves eigenstates, the susceptibility per unit volume is \cite{osti_4405425,mihaila2011lindhardfunctionddimensionalfermi}
\begin{align}\label{eq:susceptibility_volume}
    & \tilde{\chi}_V  (\mathrm{k})= \frac{1}{4}\frac{m}{2\pi^2\hbar^2\mathrm{k}}\int_0^{\infty}\mathrm{q}  n_{\mathrm{F}}^{\mathcal{V}}(\mathrm{q},\mu_V)\log{\Bigg\vert\frac{\mathrm{k} + 2\mathrm{q}}{\mathrm{k}- 2\mathrm{q }}\Bigg\vert}\; d\mathrm{q}\,.  
\end{align}

The total susceptibility of a Fermi gas is obtained by summing the susceptibility density over all volumes:
\begin{equation}\label{eq:suscept_sum_volumes}
    \chi(\mathrm{k}) = \sum_V \tilde{\chi}_V(\mathrm{k})\cdot \mathcal{V}\;, 
\end{equation}
$\mathcal{V}$ the box volume.
At zero temperature, the right-hand part of the Eq. \ref{eq:susceptibility_volume} simplifies to the Lindhard function. To calculate this quantity at finite temperature, we need to know the local chemical potential for each volume $\mu_V$, which enters the Fermi-Dirac distribution 
\begin{equation}
    n_F^{\mathcal{V}}(\mathrm{q},\mu_V) = \frac{1}{e^{\beta(\frac{\hbar^2 \mathrm{q}^2}{2m}-\mu_V)}+1}.
\end{equation}
The chemical potential in each volume is determined by matching the atom number obtained from integrating the Fermi-Dirac distribution over energy (Eq. \ref{eq:atom_number_energy}) to the local atom number set by the harmonic trap profile (Eq. \ref{eq:atom_number_harmonic_trap}). The first term is given by 
\begin{align}\label{eq:atom_number_energy}
    &N_V =\frac{\hbar^2}{m} \int_0^{\infty} \mathrm{q}\;g(\mathrm{q})n_F^{\mathcal{V}}(\mathrm{q},\mu_V)d\mathrm{q} \\
    & g(\mathrm{q}) = \frac{\mathcal{V}}{4\pi^2}\left(\frac{2m}{\hbar^2}\right)^{3/2}\sqrt{\frac{\hbar^2\mathrm{q}^2}{2m}}
\end{align}
where $g(\mathrm{q})$ is the density of states of the 3D box. The second term is obtained considering a volume centered at $\mathbf{r} = (\bar{x},\bar{y},\bar{z})$. Its atom number is given by

\begin{align}\label{eq:atom_number_harmonic_trap}
    & N_V = n(\mathbf{r} )\mathcal{V}
\end{align}
with $ n(\mathbf{r} )= - \frac{1}{\lambda_{dB}^3} \mathrm{Li}_{3/2}(-e^{\beta(\mu-U(\mathbf{r}))})$ the atomic distribution in the harmonic trap. Here, $U(\mathbf{r})$ the trap harmonic potential, $\lambda_{dB}$ the thermal deBroglie wavelength of the cloud, $\mathrm{Li}_{3/2}$ is the polylogarithm function, $\beta = 1/k_B T$. The total chemical potential $\mu$ of the harmonic trap is fixed by the total atom number.

\subsubsection{Exact harmonic oscillator eigenstates}\label{sec:exact_eigenstates}

Equation \ref{eq:def_susceptibility} can also be evaluated using the exact eigenstates of the tweezer trap, which are Hermite-Gauss functions denoted as $\Psi_{\mathbf{i}}(\mathbf{r})$. Using Wick’s theorem to evaluate the expectation values of the Fermionic operators, Eq. \ref{eq:def_susceptibility} becomes \cite{chen2023detectingfermisurfacenesting}:
\begin{align}\label{eq:susceptibility_chen}
    \chi = \sum_{\substack{\mathbf{i,j} }} & \left\vert \bra{\Psi_{\mathbf{i}}(\mathbf{r})} \cos{\mathbf{k} _{\mathrm{c}}\mathbf{r}}\cos{\mathbf{k}_{\mathrm{p}} \mathbf{r}} \ket{\Psi_\mathbf{j}(\mathbf{r})}\right\vert^2 \frac{n_F(E_{\mathbf{i}})- n_F(E_{\mathbf{j}})}{E_{\mathbf{i}}- E_{\mathbf{j}}}
\end{align}

where $\mathbf{i}= (n_x, n_y, n_z)$ denotes the $3$D harmonic oscillator states, $\Psi_{\mathbf{i}}(\mathbf{r}) = \Psi_{n_x}(x)\Psi_{n_y}(y)\Psi_{n_z}(z)$, $E_{\mathbf{i}} = E(n_x) +E(n_y)+E(n_z)$ are the eigenenergies and $n_F$ the Fermi-Dirac distribution.

This formula is fully general, and the Lindhard function for homogeneous systems is retrieved upon replacing the eignestates by plane waves and turning the sum into an integral. It suggests that the susceptibility, hence the superradiant transition can be suppressed by two mechanisms: (i) at high momentum, the energy denominator is given by the recoil, yielding the overall $1/\mathrm{E_r}$ scaling observed in Bose-Einstein condensates (see Fig.2c), and (ii) at low momentum where states are equally occupied the Fermi distributions at the numerator cancel each other, a manifestation of Pauli blocking. Note that at higher temperature the suppression can also persist in the low momentum regime when the two (thermal) distributions are very similar, which happens when the temperature is much higher than the recoil in the classical gas. 

Denoting the trap aspect ratio $r = \omega_{\perp}  / \omega_{\parallel}$, with $\omega_{\perp}, \omega_{\parallel}$ respectively the perpendicular and longitudinal trap frequencies, the susceptibility at zero temperature reads
\begin{align}\label{eq:susceptibility_eigenstates}
    \chi =&   \frac{1}{\hbar \omega_{\parallel}} \sum_{\substack{n_x,n_z \\ n_x',n_z' \\n_y}}  \frac{\Upsilon (E_F - E_{n_x, n_y, n_z}) - \Upsilon (E_F - E_{n_{x'}, n_{y}, n_{z'}})}{(n_{x'} - n_x) + r (n_{z'} - n_z)} \nonumber \\
    & \left| \int \mathrm{d} x \Psi_{n_x} (x)\Psi_{n_{x'}}(x) \cos (\mathrm{k_c}x) \right|^2 \left| \int \mathrm{d} z \Psi_{n_z} (z) \Psi_{n_{z'}}(z)  \cos (\mathrm{k_p}z) \right|^2,
\end{align}

where $E_{n_x, n_y, n_z} = \hbar \omega_{\parallel} \left( n_x + 1/2 \right)+\hbar \omega_{\perp} \left( n_y + n_z + 1 \right) $.
This expression can only be calculated at fixed Fermi energy, so different atom numbers can give the same result depending on the shell filling of the anisotropic harmonic trap. The main difficulty of the numerical evaluation of Eq. \ref{eq:susceptibility_eigenstates} is to ensure the convergence of the integrals as a function of the maximal number of states considered. We verified convergence at zero temperature by varying the maximum number of Hermite functions in each direction. For the results presented in Figure 2, having $24$ atoms in each spin component, which means $\mathrm{E_F} = 17\;\hbar\omega_{\parallel}$, we used $N_{\parallel}= 175$ states along the longitudinal direction and $N_{\perp} = N_{\parallel}/r$ for the transverse directions. 

\subsubsection{Comparison between the two simulations}
Figure \ref{fig:epsart}a shows the susceptibility calculated using LDA and exact harmonic oscillator eigenstates. Both methods reproduce the qualitative shape of the curve and capture the $1/\mathrm{k^2}$ dependence of the low trap depth regime. Quantitative differences appear at the peak, where the exact shape of the eigenstates has the biggest effect. In the high trap depth regime ($\mathrm{k/k_F}<1$), the local density approximation is not valid as the trap potential varies significantly over the length of the Fermi wavevector. The discrepancy in the asymptotic behavior at high momentum is due to the difference between the Fermi energy of the anisotropic harmonic oscillator and the LDA. LDA captures the continuous variation between the shell filling effects of the exact eigenstates calculations but neglects the finite-size effects of the trap. For instance, the largest transverse trap frequencies explored in Figure 2 represents $1.7\,\mathrm{E_r}$, approaching the Lamb-Dicke regime. 
For the parameters studied here, the exact eigenstates calculation provides the most accurate description of shell-dependent features, while LDA gives a reliable approximation for larger atom numbers and large momentum or when smoothing over closed-shell effects.

Importantly, in the grand-canonical ensemble the atom number varies in discrete steps, preventing the description of open shells. Accurate calculations in the canonical ensemble require directly manipulating Slater determinants, as done recently to study shell filling effects in light-matter interactions \cite{ortuñogonzalez2025paulicrystalsuperradiance}.

\subsubsection{Magnetic superradiance} \label{sec:magnetic_susceptibility}
In Figure 4 we consider a situation where the cavity is red--detuned from one spin state and blue--detuned from the other. Considering $g_{\uparrow}=g_{\downarrow}=g$, $\Omega_{\uparrow} = \Omega_{\downarrow}=\Omega$ and $\Delta_\downarrow = \Delta =  - \Delta_\uparrow$, this Hamiltonian can be written as 
\begin{align}
     \hat{\mathrm{H}}& = \tilde{\Delta}_{\mathrm{p}}\hat{a}^{\dagger}\hat{a} - \frac{g\Omega}{2\Delta}(\hat{\theta}_\downarrow -\hat{\theta}_{\uparrow}) (\hat{a}+ \hat{a}^{\dagger}) 
\end{align}
The cavity field couples to the operator 
\begin{align}\hat\Theta = \hat{\theta}_\downarrow- \hat{\theta}_\uparrow  =  \int d\mathbf{r} \;  \cos{(\mathbf{k_c \cdot r})}\cos{(\mathbf{k_p \cdot r})} \;(\hat{n}_{\downarrow}\;(\mathbf{r})-\hat{n}_{\uparrow}(\mathbf{r})),
\end{align}
determined by the magnetization density of the gas. In the superradiant phase, the two spin species redistribute to the maxima and minima of the interference pattern, thus creating a density modulation with a spin-wave character. 

\end{document}